\documentclass[reprint,amsmath,amssymb,aps,superscriptaddress,nofootinbib]{revtex4-1}
\usepackage{graphicx}
\usepackage{dcolumn}
\usepackage{bm}
\usepackage[utf8]{inputenc}
\usepackage{float}
\usepackage{hyperref}
\usepackage{amsmath, amsthm, amsfonts,amssymb}
\usepackage[svgnames]{xcolor}
\usepackage{mathtools}
\usepackage{caption}
\usepackage{subcaption}
\usepackage{xcolor}
\usepackage{soul}
\usepackage{xcolor}
\usepackage{appendix}
\usepackage{orcidlink}
\usepackage{booktabs}

\usepackage{comment}

\definecolor{teal}{RGB}{0, 128, 128}
\hypersetup{ colorlinks=true, linkcolor=teal, citecolor=teal, urlcolor=teal}

\begin{document}

\title{Black string immersed in perfect fluid dark matter}

\author{L. G. Barbosa \orcidlink{0009-0007-3468-3718}}
\email{leonardo.barbosa@posgrad.ufsc.br}
\affiliation{Departamento de Física, CFM - Universidade Federal de \\ Santa Catarina; C.P. 476, CEP 88.040-900, Florianópolis, SC, Brazil}

\author{L. C. N. Santos \orcidlink{0000-0002-6129-1820}}
\email{luis.santos@ufsc.br}
\affiliation{Departamento de Física, CFM - Universidade Federal de \\ Santa Catarina; C.P. 476, CEP 88.040-900, Florianópolis, SC, Brazil}

\begin{abstract}
We present an exact four-dimensional black string solution immersed in perfect fluid dark matter within an anti-de Sitter background. By solving the Einstein field equations for an anisotropic fluid, we obtain a metric function that modifies the standard black string geometry through a logarithmic term governed by the dark matter parameter $\alpha$. The event horizon radii are analytically determined using the Lambert $W$ function, and the Kretschmann scalar confirms a genuine curvature singularity at the origin alongside the expected asymptotic behavior. Furthermore, we evaluate the thermodynamic properties of the solution. The heat capacity diverges at a critical horizon radius for $\alpha>0$, a behavior commonly associated with a thermodynamic phase transition in a regime where the weak energy condition is violated.
\end{abstract}

\maketitle

\section{Introduction}\label{Introduction}

Asymptotically anti-de Sitter (AdS) spacetimes have long played a central role in gravitational physics, both because of their relevance in the context of the AdS/CFT correspondence and because they provide a natural arena for studying the interplay between geometry, thermodynamics and horizon structure. In this setting, exact solutions of Einstein's equations with nontrivial topology have been extensively investigated, including topological black holes, cylindrical geometries and extended objects whose properties differ markedly from the standard spherically symmetric case \cite{Surya:2001vj, Cardoso:2001vs, Morley:2018lwn}. These configurations are important not only as idealized gravitational systems, but also as testbeds for matter couplings, modified gravity theories and semiclassical phenomena \cite{Ali:2019mxs,Deglmann:2025mcl,Ahmed:2025sav,Barbosa:2025scy, Santos:2026bjq}.

Among these solutions, black strings occupy a distinguished place. Originally introduced by Lemos in four-dimensional AdS gravity \cite{Lemos:1994xp, Lemos:1995cm}, black strings describe cylindrically symmetric spacetimes with an event horizon extended along one spatial direction. Since then, they have been studied in several contexts, including dilaton gravity, massive gravity, mimetic gravity, Lifshitz backgrounds, rainbow gravity and asymptotically safe gravity \cite{Sa:1995vs, Hendi:2020apr, Sheykhi:2020fqf, Ghosh:2019eoo, Boonserm:2019mon, Darlla:2023qgf, Nilton:2022jji, Lessa:2024gbd, Darlla:2024bsv}. More recently, black strings have also been explored from the perspective of thermodynamics, perturbations, geodesics and analogue models, confirming that they remain a fertile ground for investigating how matter content and spacetime asymptotics shape observable and geometrical features \cite{Kumar:2023agi, Pereira:2024cfa, Ahmed:2025sav}.

The inclusion of dark matter in compact-object spacetimes has attracted considerable attention in recent years. A widely used phenomenological description is the perfect fluid dark matter (PFDM) model, which captures essential features of dark matter through an anisotropic stress-energy tensor while preserving analytical tractability \cite{Li:2012zx, Xu:2016ylr}. This framework has been applied successfully to black holes in several settings, revealing changes in horizon structure, thermodynamic stability and phase transitions \cite{Hamil:2024neq, Hamil:2024nrv}. In the case of black strings, dark matter effects have also begun to be investigated \cite{Cunha:2022kep}; however, to the best of our knowledge, an exact black string solution immersed in PFDM within an AdS background has not yet been presented in the literature. This gap motivates the present work.

Following this motivation, we construct an exact four-dimensional black string solution coupled to PFDM in the presence of a negative cosmological constant. The strategy adopted here is to solve the Einstein field equations for a cylindrically symmetric ansatz and to identify how the dark matter parameter modifies the standard Lemos geometry through a logarithmic correction. We then analyze the resulting horizon structure, curvature singularities and thermodynamic quantities, with particular emphasis on how the PFDM parameter controls the causal and thermal properties of the spacetime. In this sense, our study extends previous investigations of black strings by incorporating a matter source whose influence is both physically motivated and mathematically distinctive.

The paper is organized as follows. In Sec.~\ref{Field_equation}, we establish the theoretical framework by solving the Einstein field equations with a negative cosmological constant, obtaining the exact black string metric and explicitly showing the logarithmic modification induced by the perfect fluid dark matter parameter. In Sec.~\ref{Event_horizon_and_Kretschmann_scalar}, we analyze the geometric structure of the spacetime. We determine the event horizon radii analytically via the Lambert $W$ function, examining the distinct branches based on the sign of the dark matter parameter, and compute the Kretschmann scalar to characterize the central curvature singularity as well as the asymptotic behavior. In Sec.~\ref{Thermodynamics}, we investigate the thermodynamic properties of the solution by deriving closed-form expressions for the Hawking temperature and the heat capacity. We assess the local thermodynamic stability and analyze the divergence in the heat capacity at a critical radius. Finally, in Sec.~\ref{Discussion_and_conclusions}, we summarize our main findings and outline potential avenues for future research.

\section{Field equation}\label{Field_equation}

In this section, we obtain the solution for a static black string in the presence of a perfect fluid dark matter (PFDM). To begin, we consider Einstein's field equations with a negative cosmological constant (setting $G=c=1$):
\begin{equation}
  \mathcal{G}_{\mu\nu} = R_{\mu\nu} - \frac{1}{2} R g_{\mu\nu} - \frac{3}{\ell^{2}} g_{\mu\nu} = 8\pi T_{\mu\nu},
\end{equation}
where $R_{\mu\nu}$ is the Ricci tensor, $R$ is the scalar curvature, $g_{\mu\nu}$ is the metric tensor, $\ell$ denotes the AdS length scale, and $T_{\mu\nu}$ is the energy-momentum tensor. For the dark matter fluid, we adopt the following form \cite{Li:2012zx, Xu:2016ylr, Hamil:2024neq, Hamil:2024nrv}:
\begin{equation}
    T_{\mu}^{\nu} = \operatorname{diag}\left(-\rho, -\rho, (\epsilon-1)\rho, (\epsilon-1)\rho\right). 
\end{equation}

Since our goal is to describe black string configurations, we consider a stationary and axisymmetric line element adapted to cylindrical symmetry:
\begin{equation}
    ds^{2} = -f(r)\,dt^{2} + \frac{dr^{2}}{f(r)} + r^{2}\,d\varphi^{2} + \frac{r^{2}}{\ell^{2}}\,dz^{2},
\end{equation}
with the coordinates ranging as $-\infty < t < +\infty$, $0 \leq r < +\infty$, $0 \leq \varphi < 2\pi$, and $-\infty < z < +\infty$. The function $f(r)$ is to be determined by the field equations. From this ansatz, the nonvanishing components of the Einstein tensor $\mathcal{G}_{\mu\nu}$ take the form:
\begin{align}
    \mathcal{G}_{tt} &= -f(r)\left(\frac{f'(r)}{r} + \frac{f(r)}{r^{2}} - \frac{3}{\ell^{2}}\right), \\
    \mathcal{G}_{rr} &= \frac{1}{f(r)}\left(\frac{f'(r)}{r} + \frac{f(r)}{r^{2}} - \frac{3}{\ell^{2}}\right), \\
    \mathcal{G}_{\phi\phi} &= r^{2}\left(\frac{1}{2}f''(r) + \frac{f'(r)}{r} - \frac{3}{\ell^{2}}\right), \\
    \mathcal{G}_{zz} &= \frac{1}{\ell^{2}}\,\mathcal{G}_{\phi\phi}.
\end{align}

Having characterized the geometric sector, we now turn to the matter content. The components of the energy-momentum tensor corresponding to the perfect fluid dark matter are given by:
\begin{align}
    8\pi T_{tt} &= 8\pi\rho\,f(r), \\
    8\pi T_{rr} &= -\frac{8\pi\rho}{f(r)}, \\
    8\pi T_{\phi\phi} &= 8\pi\rho\,r^{2}(\epsilon-1), \\
    8\pi T_{zz} &= \frac{8\pi}{\ell^{2}}\,\rho\,r^{2}(\epsilon-1).
\end{align}

Equating the appropriate components of the Einstein field equations, we obtain the following two independent equations:
\begin{align}
     \frac{1}{r}f'(r) + \frac{1}{r^{2}}f(r) - \frac{3}{\ell^{2}} &= -8\pi\rho,\\
      \frac{1}{2}f''(r) + \frac{1}{r}f'(r) - \frac{3}{\ell^{2}} &= 8\pi\rho(\epsilon-1).
\end{align}

Combining these expressions, we can eliminate the density $\rho$ and arrive at a single differential equation for $f(r)$:
\begin{equation}
    \frac{1}{2}f''(r) + \frac{\epsilon}{r}f'(r) + (\epsilon-1)\frac{1}{r^{2}}f(r) - \frac{3\epsilon}{\ell^{2}} = 0.
\end{equation}
To solve this equation, we adopt an ansatz that separates the AdS background from a perturbation induced by the matter source:
\begin{equation}
    f(r) = \frac{r^{2}}{\ell^{2}} - U(r).
\end{equation}

Substituting this form into the differential equation yields the same equation found in \cite{Li:2012zx}:
\begin{equation}
    r^{2}U''(r) + 2\epsilon\,r\,U'(r) + 2(\epsilon-1)U(r) = 0.
\end{equation}

The general solution of this Euler-type equation depends on the value of the parameter $\epsilon$:
\begin{equation}
    U(r) = \begin{cases}
        \dfrac{2M}{r} - \dfrac{r^{2(1-\epsilon)}}{r_{\epsilon}}, & \epsilon \neq \dfrac{3}{2}, \\[1.2em]
        \dfrac{2M}{r} - \dfrac{\alpha}{r}\ln\left(\dfrac{r}{|\alpha|}\right), & \epsilon = \dfrac{3}{2}.
    \end{cases}
\end{equation}

Where $\alpha$ is an integration constant. In what follows we focus on the case $\epsilon = \frac{3}{2}$, which leads to the metric function:
\begin{equation}
    f(r) = \frac{r^{2}}{\ell^{2}} - \frac{2M}{r} + \frac{\alpha}{r}\ln\left(\frac{r}{|\alpha|}\right).
\end{equation}

Finally, by inserting this expression for $f(r)$ back into the field equations, we can explicitly compute the energy density and pressure of the dark matter fluid. They are found to be:
\begin{align}
     \rho &= -\frac{\alpha}{8\pi r^{3}},\\
     p &= -\frac{\alpha}{16\pi r^{3}},
\end{align}
where $p_z=p_{\phi} \equiv p$. Notice that the energy density and pressure satisfy the relation $p = \rho/2$, which characterizes the equation of state for the perfect fluid dark matter in this scenario. The sign of the integration constant $\alpha$ plays a crucial role in determining the physical nature of the fluid: for $\alpha > 0$, both $\rho$ and $p$ are negative, indicating a violation of the weak energy condition, a behavior typically associated with exotic matter or dark energy-like components. Conversely, if $\alpha < 0$, the density and pressure become positive, corresponding to a more conventional matter source. 

\section{Event horizon and Kretschmann scalar}\label{Event_horizon_and_Kretschmann_scalar}

In this section, we examine the event horizon structure and the behavior of curvature singularities using the Kretschmann scalar. The event horizon is determined by the condition $f(r) = 0$, which, for the metric function, leads to the following transcendental equation:
\begin{equation}
   r^{3} + \alpha\ell^{2}\ln\left(\frac{r}{\left|\alpha\right|}\right) - 2M\ell^{2} = 0.
\end{equation}

Since $\alpha$ is an integration constant, its sign plays a fundamental role in the horizon structure, and we must analyze the two cases separately.

For $\alpha > 0$, the horizon radius can be expressed in terms of the principal branch of the Lambert $W$ function as:
\begin{equation}
    r_{+} = \left[\frac{\alpha\ell^{2}}{3}\right]^{\frac{1}{3}}\left[W_{0}\left(\frac{3\alpha^{2}}{\ell^{2}}e^{\frac{6M}{\alpha}}\right)\right]^{\frac{1}{3}}.
\end{equation}

For $\alpha < 0$, it is convenient to write $\alpha = -|\alpha|$. In this case, the horizon equation admits two possible branches, corresponding to the $k=0$ and $k=1$ branches of the Lambert $W$ function:
\begin{equation}
    r_{k} = \left[\frac{|\alpha|\ell^{2}}{3}\right]^{\frac{1}{3}}\left[-W_{k}\left(-\frac{3\alpha^{2}}{\ell^{2}}e^{-\frac{6M}{|\alpha|}}\right)\right]^{\frac{1}{3}}, 
\end{equation}
subject to the existence condition
\begin{equation}
    \frac{3\alpha^{2}}{\ell^{2}}e^{-\frac{6M}{|\alpha|}} \leq \frac{1}{e}.
\end{equation}
Depending on the parameter values, these solutions may correspond to distinct horizon radii, with the larger one representing the event horizon.

Now, we will analyze the curvature singularities using the Kretschmann scalar $K = R_{\mu\nu\rho\sigma} R^{\mu\nu\rho\sigma}$. For the black string metric obtained previously, a direct calculation results in:
\begin{align}
    K &= \frac{24}{\ell^{4}} + \frac{4\alpha}{\ell^{2}r^{3}} \nonumber 
    \\ &\quad + \frac{13\alpha^{2}+40\alpha M+48M^{2}}{r^{6}} \nonumber 
    \\ &\quad + \frac{4\alpha}{r^{6}}\ln\left(\frac{r}{\left|\alpha\right|}\right)\left[3\alpha\ln\left(\frac{r}{\left|\alpha\right|}\right) - 5\alpha - 12M\right].
\end{align}

This expression reveals the asymptotic behavior of the spacetime: as $r \to \infty$, $K \to 24/\ell^{4}$, confirming that the geometry approaches pure AdS space. In the opposite limit, $r \to 0$, the terms proportional to $r^{-6}$ dominate, indicating a genuine curvature singularity at the origin, which is hidden behind the event horizon whenever the latter exists. 

\section{Thermodynamics}\label{Thermodynamics}

We now discuss the thermodynamic properties of the black string solution, with particular emphasis on the Hawking temperature and heat capacity. In the standard framework of black hole thermodynamics, the Hawking temperature associated with the event horizon is given by \cite{Hawking:1975vcx,Bekenstein:1973ur}
\begin{equation}
    T = \left. \frac{f'(r)}{4\pi} \right|_{r=r_{+}},
\end{equation}
where \(r_{+}\) denotes the radius of the event horizon, defined as the largest root of \(f(r)=0\). Evaluating the derivative of the metric function at the horizon, we obtain
\begin{equation}
    T = \frac{1}{4\pi} \left( \frac{2r_{+}}{\ell^{2}} + \frac{1}{r_{+}^{2}} \left[ 2M + \alpha - \alpha \ln \left( \frac{r_{+}}{|\alpha|} \right) \right] \right).
\end{equation}

The parameter $M$ can be expressed in terms of the horizon radius by imposing the horizon condition $f(r_{+})=0$, which gives
\begin{equation}
    M = \frac{r_{+}}{2} \left[ \frac{r_{+}^{2}}{\ell^{2}} + \frac{\alpha}{r_{+}} \ln \left( \frac{r_{+}}{|\alpha|} \right) \right].
\end{equation}

Substituting this expression for the mass parameter into the Hawking temperature and simplifying, we obtain a remarkably simple form
\begin{equation}
    T = \frac{r_{+}}{4\pi} \left( \frac{3}{\ell^{2}} + \frac{\alpha}{r_{+}^{3}} \right).
\end{equation}

We now examine the local thermodynamic stability of the black string, which is governed by the sign of the heat capacity. A positive heat capacity indicates local thermodynamic stability, while a negative value signals instability. The heat capacity is defined as
\begin{equation}
    C = \left( \frac{dM}{dT} \right)_{r=r_{+}}.
\end{equation}

Computing the derivatives of the mass and temperature with respect to the horizon radius, we obtain
\begin{equation}
 C=\frac{2\pi r_{+}^{2}\left(r_{+}^{3}+\frac{\alpha\ell^{2}}{3}\right)}{\left(r_{+}^{3}-\frac{2\alpha\ell^{2}}{3}\right)}.
\end{equation}

The heat capacity exhibits a divergence at a critical horizon radius $r_{+}^{c}$, signaling a phase transition in the thermodynamic behavior of the black string. This critical radius is determined by the vanishing of the denominator in the expression above, which yields the condition $  r_{+}^{c} = \left( 2 \alpha \ell^{2}/3 \right)^{\frac{1}{3}}$. For a real and positive critical radius to exist, we require $\alpha \ell^{2} \geq 0$. Since $\ell^{2} > 0$, this condition implies $\alpha \geq 0$. Notably, from our earlier analysis of the energy-momentum tensor, we found that $\alpha > 0$ corresponds to negative energy density and pressure, indicating a violation of the weak energy condition. Thus, the existence of a critical radius and the associated phase transition are directly linked to the exotic nature of the perfect fluid dark matter.

\begin{figure}[H]
    \centering
    \includegraphics[width=0.9\linewidth]{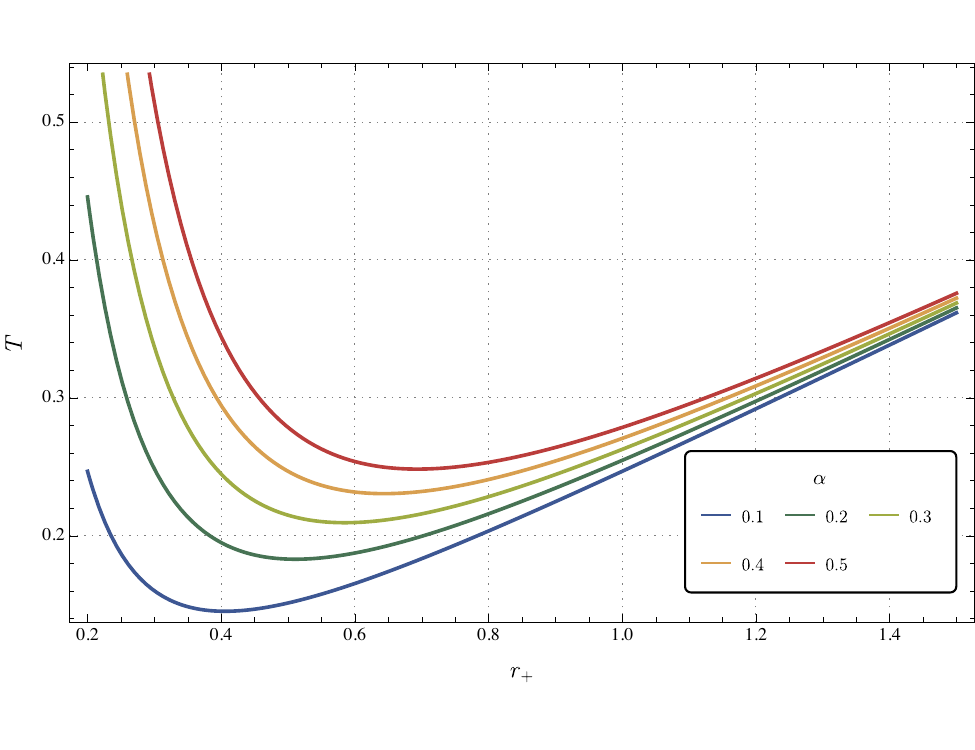}
    \caption{Hawking temperature $T$ versus event horizon radius $r_+$ for selected values of $\alpha>0$.}
    \label{fig:hawking_tempI}
\end{figure}

\begin{figure}[H]
    \centering
    \includegraphics[width=0.9\linewidth]{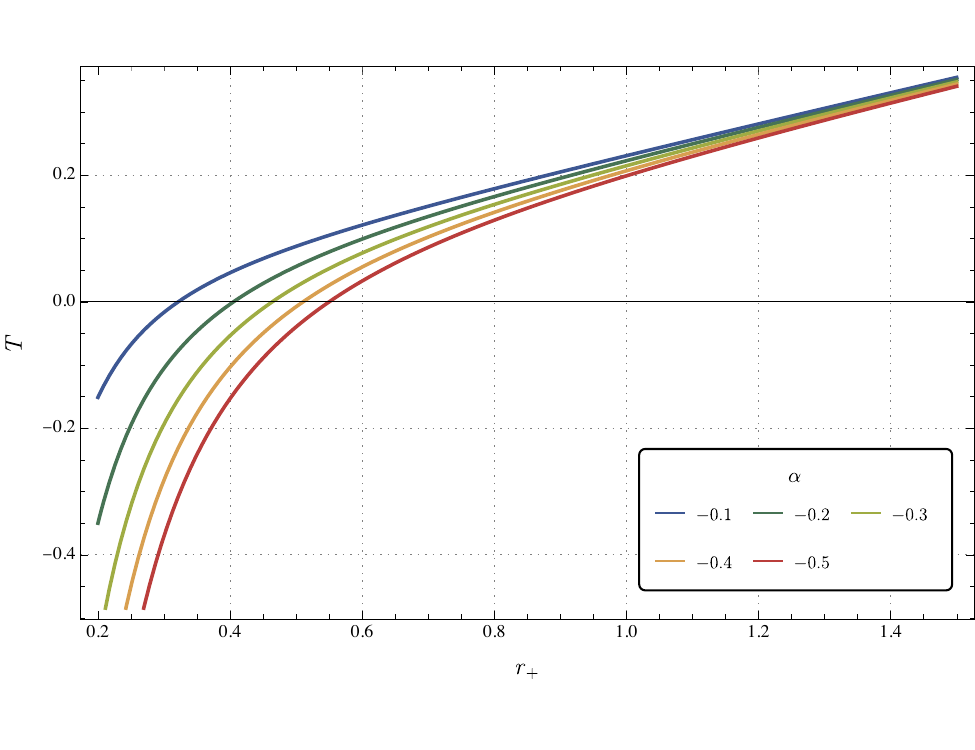}
    \caption{Hawking temperature $T$ versus event horizon radius $r_+$ for selected values of $\alpha<0$.}
    \label{fig:hawking_tempII}
\end{figure}

\begin{figure}[H]
    \centering
    \includegraphics[width=0.9\linewidth]{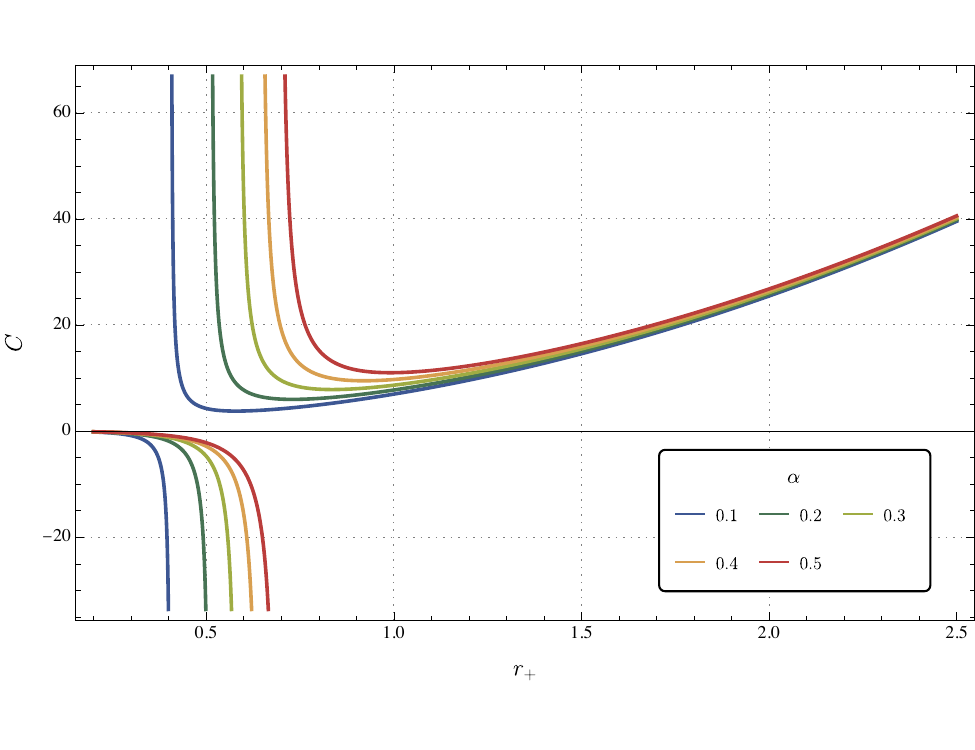}
    \caption{Heat capacity $C$ versus event horizon radius $r_+$ for selected values of $\alpha>0$.}
    \label{fig:heat_capacityI}
\end{figure}

\begin{figure}[H]
    \centering
    \includegraphics[width=0.9\linewidth]{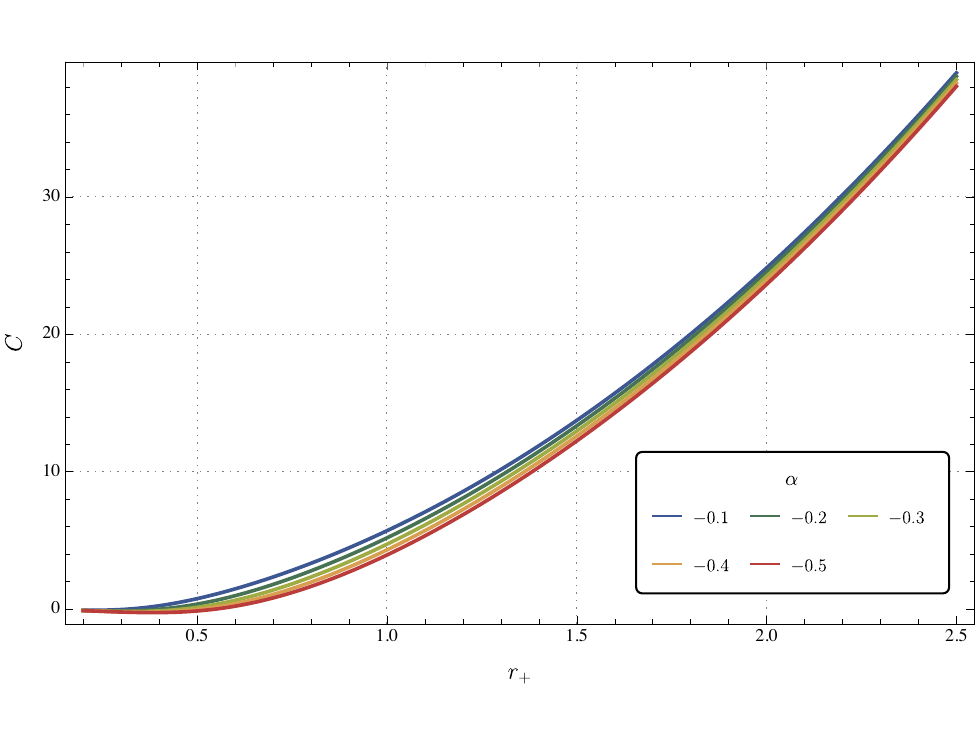}
    \caption{Heat capacity $C$ versus event horizon radius $r_+$ for selected values of $\alpha<0$.}
    \label{fig:heat_capacityII}
\end{figure}

To analyze local thermodynamic stability, we plot the Hawking temperature $T$ and heat capacity $C$ versus the event horizon radius $r_+$ (setting $\ell=1$). For $\alpha > 0$ (Figs.~\ref{fig:hawking_tempI} and \ref{fig:heat_capacityI}), each temperature curve exhibits a minimum at $r_{+}^{c} = (2\alpha\ell^2/3)^{1/3}$, which exactly coincides with the point where the heat capacity diverges. In this case, for $r_+ < r_{+}^{c}$, we have $C < 0$, while for $r_+ > r_{+}^{c}$, $C > 0$. This behavior indicates a thermodynamic phase transition and a change in local stability, with larger values of $\alpha$ shifting the critical radius $r_{+}^{c}$ to larger radii.

Conversely, for $\alpha < 0$ (Figs.~\ref{fig:hawking_tempII} and \ref{fig:heat_capacityII}), the thermodynamic behavior changes due to the sign of the parameter. Evaluating the heat capacity equation for negative values of $\alpha$ reveals that its denominator remains strictly positive for all $r_+ > 0$. Consequently, the heat capacity does not diverge, indicating the absence of the phase transition observed in the $\alpha > 0$ case.  For event horizons larger than this value, both $T$ and $C$ are positive, showing that the black string is locally thermodynamically stable in this regime.

\section{Discussion and conclusions}\label{Discussion_and_conclusions}

In this work, we presented an exact four-dimensional black string solution in an anti-de Sitter (AdS) background surrounded by a perfect fluid dark matter (PFDM) distribution. Starting from a static and cylindrically symmetric ansatz, we solved the Einstein field equations for an anisotropic fluid in the presence of a negative cosmological constant. The resulting metric function modifies the standard Lemos black string geometry by introducing a logarithmic term, which is governed by the dark matter parameter $\alpha$.

We analyzed the geometric structure of this spacetime in detail. The condition for the event horizon results in a transcendental equation, and we demonstrated that the horizon radii can be determined analytically using the Lambert $W$ function. Depending on the sign of the parameter $\alpha$, the solutions present distinct mathematical branches. Furthermore, the evaluation of the Kretschmann scalar confirmed the existence of a central curvature singularity at the origin. This analysis also verified that the spacetime asymptotically approaches the pure AdS geometry at large radial distances.

The thermodynamic properties of the black string were investigated by calculating closed-form expressions for the Hawking temperature and the heat capacity. We found that the thermal behavior of the solution depends directly on the sign of the dark matter parameter. For $\alpha > 0$, which corresponds to a fluid with negative energy density and pressure that violates the weak energy condition, the heat capacity diverges at a specific critical horizon radius. This divergence indicates a thermodynamic phase transition, separating a region of local instability for smaller horizons from a region of stability for larger horizons.

Conversely, for the $\alpha < 0$ case, the mathematical behavior of the thermodynamic quantities is different. The denominator of the heat capacity remains positive, meaning there are no divergences or phase transitions. In this scenario, the system admits an extremal black string configuration with a zero-temperature state at a minimum physical radius. For all physically valid event horizons larger than this extremal limit, the Hawking temperature increases monotonically, and the heat capacity is strictly positive. This establishes that the black string is locally thermodynamically stable in this regime.

In conclusion, the inclusion of perfect fluid dark matter modifies the geometric configuration and the thermal properties of AdS black strings. These results provide a foundation for further research. Future work could consider extending the present exact solution by incorporating rotation or an electric charge, as well as studying the dynamic stability of the spacetime under scalar and gravitational perturbations.

\section{Acknowledgements}\label{Acknowledgements}
L.G.B would like to thank CAPES (Process number: 88887.968290/2024-00) for the financial support. LCNS would like to thank Conselho Nacional de Desenvolvimento Científico e Tecnológico - Brazil (CNPq) for financial support under Research Project No. 443769/2024-9 and Research Fellowship No. 314815/2025-2.

\bibliographystyle{unsrturl}
\bibliography{sample}

\end{document}